\documentstyle[12pt]{article}
\newcommand\beq{\begin{equation}}
\newcommand\eeq{\end{equation}}
\newcommand\bea{\begin{eqnarray}}
\newcommand\eea{\end{eqnarray}}

\begin{document}
\vspace{-2.0cm}
\bigskip

\centerline{\Large \bf The Physical Hilbert Space of SU(2) Lattice Gauge Theory} 
\vskip .8 true cm

\begin{center} 
{\bf Manu Mathur}\footnote{E. Mail: manu@bose.res.in}  

S. N. Bose National Centre for Basic Sciences \\
 JD Block, Sector III, Salt Lake City,  Calcutta 98, India.

\end{center} 
\bigskip

\centerline{\bf Abstract}

We solve the Gauss law of SU(2) lattice gauge theory using the  harmonic 
oscillator prepotential formulation. We construct a generating function 
of a  manifestly gauge invariant and orthonormal basis in the physical 
Hilbert space of (d+1) dimensional SU(2) lattice gauge theory. The 
resulting orthonormal physical states 
are given in closed form. The generalization to SU(N) gauge group is discussed. 

\vskip .4 true cm

\section{\bf Introduction}

The physical states of gauge theories are gauge invariant. Therefore, 
an important problem is to label and construct a complete orthonormal 
basis in the physical (gauge invariant) Hilbert space of gauge theories. 
The motivation of the present work is to address this problem within the 
framework of lattice gauge theories using the recently proposed harmonic 
oscillator prepotential formalism.  We will be working with the Hamiltonian 
formulation of (d+1) dimensional lattice gauge theories.  Further, to keep 
the discussions simple, we will deal with SU(2) gauge groups and the generalization 
to SU(N) will be discussed at the end. 

\noindent There have been two approaches to the above problem: 
\begin{itemize}
\item The set of all Wilson loops \cite{wilson,kogut}  generates a basis in 
the physical Hilbert space which is manifestly gauge invariant. However,  
this basis is  not orthogonal and it is overcomplete \cite{mandelstam}. 
\item  Characterize the states associated with each lattice link by 
the eigenvalues of a complete set of commuting operators and then impose the Gauss 
law constraints on these states over the lattice \cite{sharat}. 
Such solutions of Gauss law, have been written  down in terms of Wigner D 
matrices in different spin representations \cite{robson}. Unlike the Wilson loop 
basis, they provide a complete orthonormal basis. However, these constructions lack 
the elegance of Wilson loop construction and are limited by the problem of 
rapid proliferation of the SU(2) group indices and the Clebsch Gordan coefficients. 
\end{itemize} 
Recently, we proposed a re-formulation of SU(2) lattice gauge 
theories in terms of harmonic oscillator prepotentials \cite{manu}. 
Two of the novel features of this formulation which are relevant for the 
present work are: 
\begin{enumerate}
\item In terms of harmonic oscillator prepotentials, the Hamiltonian 
has $SU(2) \otimes U(1)$ local gauge invariance. 
\item The prepotentials with U(1) charges transform as doublets 
under the SU(2) gauge transformations.  This enabled us to study the
gauge invariance of the theory more critically. 
\end{enumerate}
Using this formulation, we now construct a basis in the 
physical Hilbert space which has the following desired properties: 
\begin{itemize} 
\item It is orthonormal and complete.  
\item Like Wilson loops, it is manifestly gauge invariant and is given in  
a closed form in terms of the harmonic oscillator prepotentials. 
\item In d space dimensions, it is characterized by 3(d-1) quantum numbers 
per lattice site which is the number of physical transverse degrees of 
freedom of the SU(2) gluons \cite{sharat}. 
\end{itemize} 
We should emphasize that this formulation 
completely bypasses the problem of proliferation of indices leading to cumbersome 
expressions faced by the earlier approaches.  The reason for this simplification 
is the novel $SU(2) \otimes U(1)$ gauge group and the corresponding gauge 
transformation properties of the prepotentials (see section 3). These features 
of prepotential formulation of lattice gauge theories will be emphasized and 
compared with the standard formulation as we proceed.   

\noindent The plan of the paper is as follows: In section (2), we start with a 
brief introduction to SU(2) prepotential Hamiltonian formulation \cite{manu}. 
This section is included for the  the sake of completeness and for setting up 
the notations which are more suitable for the present work. 
The Section (3) is devoted to the study of physical Hilbert space in terms 
of prepotential operators. In this section, we explicitly construct a manifestly 
SU(2) gauge invariant orthonormal basis in the physical Hilbert space and 
write down the corresponding generating function. 

\section{\bf The Hamiltonian Prepotenial Formulation}

\noindent We start with SU(2) lattice gauge theory in (d+1) 
dimensions. The Hamiltonian is \cite{kogut}:  
\bea 
H = \sum_{n,i} tr E(n,i)^{2} 
+ K \sum_{plaquettes} tr \Big(U_{plaquette} + U^{\dagger}_{plaquette}\Big).    
\label{ham}   
\eea 
where, 
\bea 
U_{plaquette} = U(n,i)U(n+i,j)U^{\dagger}(n+j,i)U^{\dagger}(n,j); 
~~ E(n,i) \equiv  E^{a}(n,i){\sigma^{a} \over 2}. 
\nonumber 
\eea 
and K is the coupling constant. The index n labels the site of a d-dimensional 
spatial lattice and i,j (=1,2,...d) denote the direction of the links. Each link (n,i) 
is associated with a symmetric top, whose configuration (i.e the rotation matrix 
from space fixed to body fixed frame) is given by the operator valued SU(2) 
matrix U(n,i). The angular momenta with respect to body and space fixed frames are 
given by $E^{a}(n,i) (a=1,2,3)$ and $e^{a}(n+i,i)$ respectively. The quantization rules 
\cite{kogut} are\footnote{Note that compared to \cite{manu}, we have introduced a -ve sign infront 
of $E^{a}(n,i)$ for later convenience}: 
\bea 
\left[E^{a}(n,i),E^{b}(n,i)\right] =  i \epsilon^{abc} E^{c}(n,i),  ~~~~
\left[e^{a}(n,i),e^{b}(n,i)\right] = i \epsilon^{abc} e^{c}(n,i). 
\label{cr1} 
\eea    
$E^{a}(n,i)$ and $e^{a}(n+i,i)$ commute with each other and their magnitudes are same: 
\bea
\sum_{a=1}^{3} E^{a}(ni)E^{a}(ni) = \sum_{a=1}^{3} e^{a}(n+i,i)e^{a}(n+i,i).  
\label{cons} 
\eea 
However, the axis of rotation can have arbitrary inclinations in the two frames. Therefore, 
the complete set of commuting operators on any link (ni) are: $\sum_{a=1}^{3} E^{a}(n,i)E^{a}(n,i) 
(= \sum_{a=1}^{3} e^{a}(n+i,i)e^{a}(n+i,i)); E^{3}(n,i); e^{3}(n+i,i)$. The corresponding 
eigenvectors will be denoted by $\vert j(ni), m(ni), \bar{m}(n+i,i) \rangle$. 
 The Hamiltonian (\ref{ham}) and the quantization rules (\ref{cr1}) are invariant under: 
\bea
 E(ni) \rightarrow V(n) E(ni) V^{\dagger}(n); ~  e(ni) \rightarrow V(n) e(ni) V^{\dagger}(n); 
~ U(ni) \rightarrow V(n)U(ni)V^{\dagger}(n+i). 
\label{gt1} 
\eea 
The left and right gauge transformation on U(ni) are generated by the $E^{a}(ni)$ and 
$e^{a}(n+i,i)$ respectively. Therefore,  we should think of $E^{a}(ni)$ and  
$e^{a}(n+i,i)$ as the operators attached to the two ends (n) and (n+i) of the link (ni) 
respectively. The SU(2) Gauss law at every site (n) is: 
\bea 
\sum_{i=1}^{d} \big(E^{a}(n,i) + e^{a}(n,i)\big) = 0. 
\label{gl} 
\eea
Having fixed the notations, we now define the SU(2) prepotentials. 
To every link (n,i), we associate two doublets of harmonic oscillators 
$(a_{\alpha}(n,i), a^{\dagger}_{\alpha}(n,i))$ and $(b_{\alpha}(n+i,i), 
b^{\dagger}_{\alpha}(n+i,i))$ attached to the two ends (n) and (n+i) respectively.   
They satisfy: 
\bea
[a_{\alpha}, a^{\dagger}_{\beta}] = \delta_{\alpha,\beta};~~ [b_{\alpha}, b^{\dagger}_{\beta}]
= \delta_{\alpha,\beta} ~~~ \alpha,\beta =1,2.
\label{ho}
\eea
Using  the Jordan-Schwinger boson representation of SU(2) Lie algebra \cite{schwinger}, 
we write: 
\bea
E^{a}(n,i) \equiv a^{\dagger}(n,i){\sigma^{a} \over 2} a(n,i); ~~~
e^{a}(n,i) \equiv b^{\dagger}(n,i){{\sigma}^{a} \over 2} b(n,i). 
\label{sb}
\eea
The gauge transformation properties of electric fields (\ref{gt1}) imply that
\bea
a^{\dagger}_{\alpha}(n,i) \rightarrow V(n)_{\alpha\beta} a^{\dagger}_{\beta}(n,i), ~~~
b^{\dagger}_{\alpha}(n,i) \rightarrow V(n)_{\alpha\beta} b^{\dagger}_{\beta}(n,i). 
\label{gt3}
\eea
\noindent Thus, under SU(2) gauge transformations both a and b type prepotentials 
transform as doublets. Therefore, the manifestly SU(2) gauge invariant operators 
are: 
\bea
a^{\dagger}.b \equiv \sum_{\alpha=1}^{2} a^{\dagger}_{\alpha}b_{\alpha},~~~~ 
a^{\dagger}.\tilde{b}^{\dagger} \equiv \sum_{\alpha=1}^{2} \sum_{\beta=1}^{2} 
\epsilon_{\alpha\beta} a^{\dagger}_{\alpha} {b}^{\dagger}_{\beta} =   
a^{\dagger}_{1}b^{\dagger}_{2} - a^{\dagger}_{2}b^{\dagger}_{1}.
\label{mgi} 
\eea
The constraint (\ref{cons}) implies that the occupation numbers 
of the harmonic oscillator prepotentials located at (n) and (n+i) ends 
of the link (n,i) are equal.  
\bea
a^{\dagger}(n,i).a(n,i) = b^{\dagger}(n+i,i).b(n+i,i)  \equiv N(n,i);~~i=1,2,...,d.  
\label{consho} 
\eea 
Therefore, the Hilbert space  of pure SU(2) lattice gauge 
theory is  characterized by the following orthonormal state vectors 
at each link: 
\bea 
\vert j(n,i),m(n,i),\bar{m}(n+i,i) \rangle = 
{(a^{\dagger}_1)^{j+m} \over \sqrt{(j+m)!}} 
{(a^{\dagger}_2)^{j-m} \over \sqrt{(j-m)!}} 
{(b^{\dagger}_1)^{j+\bar{m}} \over \sqrt{(j+\bar{m})!}}  
{(b^{\dagger}_2)^{j-\bar{m}} \over \sqrt{(j-\bar{m})!}} 
| {}^{0~~0}_{0~~0} \rangle . 
\label{hs}
\eea 
In (\ref{hs}) $a^{\dagger}$ and $b^{\dagger}$ (also m and $\tilde{m}$) are defined 
at the two ends  n and n+i of the link  (n,i) respectively.  
From now onwards we will denote the Hilbert space consisting of all possible 
state vectors (\ref{hs}) at all the links of the lattice as $\tilde{\cal H}$. 
Note that $\tilde{\cal H}$
defines the dual Hilbert space as it is characterized by the eigenvalues values of the angular momentum 
operators.  The defining equations (\ref{sb}) for the SU(2) prepotentials and 
the Hilbert space in (\ref{hs}) are invariant under: 
\bea 
a^{\dagger}_{\alpha}(n,i) \rightarrow expi\theta(n,i)~ a^{\dagger}_{\alpha}(n,i); ~~  
b^{\dagger}_{\alpha}(n+i,i) \rightarrow exp-i\theta(n,i) ~b^{\dagger}_{\alpha}(n+i,i).   
\label{u1}
\eea 
In (\ref{u1}), $\theta(n,i)$ is a phase angles at link (n,i). This is a novel U(1) local 
gauge invariance. Note that this is {\it not an abelian subgroup} of the SU(2) gauge group. 
The constraint (\ref{consho}) now becomes the Gauss law for this abelian gauge invariance. 
The $SU(2) \otimes U(1)$ gauge invariance imply \cite{manu}:
\bea
U(n,i)_{\alpha\beta} = F(n,i)  \big(a^{\dagger}_{\alpha}(n,i) 
\tilde{b}^{\dagger}_{\beta}(n+i,i)
+ \tilde{a}_{\alpha}(n,i) {b}_{\beta}(n+i,i)\big) F(n,i).
\label{dh}
\eea
In (\ref{dh}), $F(n,i) \equiv  {1 \over \sqrt{N(n,i)+1}}$ with N(n,i) defined in (\ref{consho}) is the 
normalization factor. 

\section{A Manifestly SU(2) Gauge Invariant Basis} 

In this section, we  construct the generating function which produces a manifestly 
SU(2) gauge invariant orthonormal basis.  The  abelian gauge invariance 
(\ref{u1}) and the associated Gauss law (\ref{consho}) are  simple and will be 
incorporated at the end.  
Before we start the construction of the gauge invariant basis, we would like to emphasize 
the following two features of the prepotential formulation which are relevant for 
the construction of generating function in this section. 

\begin{enumerate} 
\item In terms of the link variables U(n,i), the state given in (\ref{hs}): 
\bea 
|j(n,i),m(n,i),\bar{m}(n,i)\rangle = \sum_{i_1,....i_{2j} \in S_{2j}} U_{m_{i_1}\bar{m}_1}
U_{m_{i_2} \bar{m}_2}...U_{ m_{i_{2j}} \bar{m}_{2j} }|0 \rangle.
\label{st}
\eea 
In (\ref{st}), $(m_1,m_2,..,m_{2j})$ and $(\bar{m}_1,\bar{m}_2,..,\bar{m}_{2j})$ 
are the two sets of  $\pm {1 \over 2}$  with constraints:  
$m_1+m_2 +..m_{2j} = m$, $\bar{m}_1+\bar{m}_2 +..\bar{m}_{2j} = 
\bar{m}$ and $S_{2j}$ is the permutation group of order 2j.  The construction (\ref{st}) 
though the link operators (unlike (\ref{hs})) becomes 
more and more complicated as j increases.  Thus, the  characterization of the  
Hilbert space, even before imposing the Gauss law constraints, through the prepotential 
formulation is much simpler than the standard formulation. 
\item The dynamical variables at site n are $a^{\dagger}(n,i)$ and $b^{\dagger}(n,i)$, 
i=1,2,...,d. 
Under SU(2) gauge transformation at site n, they all transform as SU(2) doublets like 
matter fields. Therefore, the problem of constructing the SU(2) gauge invariant 
Hilbert space over the entire lattice using the  link operators U reduces to a 
(local) problem of constructing  SU(2) singlets out of SU(2) harmonic oscillator 
doublets at one lattice site. The additional U(1) Gauss law then connects these 
local SU(2) gauge invariant Hilbert spaces at different lattice sites. 
\end{enumerate} 
The above discussion suggests that we collect all the 2d prepotential doublets meeting 
at a particular lattice site n and label them\footnote{More explicitly, $a^{\dagger}_{\alpha}[i]  \equiv
a^{\dagger}_{\alpha}(n,i)$;  and $a^{\dagger}_{\alpha}[d+i]  \equiv b^{\dagger}_{\alpha}(n,i)$; 
i=1,2,.....,d. From now on, our analysis will be only at a particular site n. Therefore, we will 
suppress the site index and show it explicitly only when required.} as $a_{\alpha}[i], 
a^{\dagger}_{\alpha}[i]$; 
i=1,2,..,2d. The corresponding angular momentum operators are $ a^{\dagger}[i] 
{\sigma^{a} \over 2} a[i] $ and will be denoted by $J^{a}[i]$. The  
eigenvalues of $\sum_{a=1}^{3} J^{a}[i]J^{a}[i]$ and $J^{(a=3)}[i]$ will be denoted by 
$j_{i}$ and $m_{i}$ respectively.  Therefore, a generic  state in the 
Hilbert space ${\tilde{ \cal{H}}}_{n}$ living at the site n is labelled by 4d quantum 
numbers: 
\bea 
\prod_{i=1}^{2d} \otimes \vert j_{i}, m_{i} \rangle. 
\label{dps} 
\eea   
The complete Hilbert space ${\tilde{\cal H}}$ consisting of state vectors in  (\ref{hs}) is 
obtained by taking direct product: 
\bea 
{\tilde{\cal H}}  = \prod_{n}{}^{\prime} \otimes {\tilde{\cal H}}_{n}.
\label{physs} 
\eea 
In (\ref{physs}), the prime over the direct product over all lattice sites implies that the 
direct product is taken such that U(1) Gauss law (\ref{consho}) is satisfied. 

\noindent The SU(2) Gauss law (\ref{gl}) at site n now takes a simpler form: 
\bea 
J^{a}_{total} = J^{a}[1]+J^{a}[2]+.........+J^{a}[2d] = 0. 
\label{glf} 
\eea 
It  simply states that the sum of all the 2 d angular momenta 
meeting at a site (n) is zero. The new form of SU(2) Gauss law (\ref{glf}) 
immediately implies that the physical Hilbert space 
living at site n and denoted by $\tilde{\cal H}_{n}^{p}$ can be trivially 
characterized as \cite{manu}: 
\bea
|\vec{l}> \equiv \left\vert \begin{array}{cccc} 
l_{12} & l_{13}  & ... & l_{1~2d}     \\
 & l_{23} & ... & l_{2~2d} \\
 &  & . & . \\
 &  &  & ~~~~~l_{2d-1~2d} \\
\end{array} \right \rangle =  \prod_{{}^{{i},{j}}_{{j} > 
{i}}} \left(a^{\dagger}[i].\tilde{a}^{\dagger}[j]\right)^{l_{{i}\bar{j}}}
|0> . 
\label{givv} 
\eea 
In (\ref{givv}), $l_{ij} \left(\equiv l_{ji}\right)$ are $N_{d} = d(2d-1)$ 
 +ve integers at site n which are invariant under the SU(2) gauge transformations. 
The states (\ref{givv})  $ \in \tilde{\cal H}_{n}^{p}$ with total angular 
momentum zero are also the eigenvectors of J[i].J[i], i= 1,2,..,2d 
with eigenvalues $j_{i}(j_{i}+1)$ where  $2 j_{i} = \left(\sum_{{j} \neq {i}} 
l_{{i}{j}}\right)$. As the above 2d+2 mutually commuting operators do not form 
the  complete set of commuting operators (except in the trivial case of d=1), 
the SU(2) invariant basis (\ref{givv}), like the standard basis obtained by all 
Wilson loops, is obviously overcomplete and not 
orthonormal. Infact, (\ref{givv}) gives the dual description of the basis 
obtained by all possible Wilson loops \cite{manu}.    

\noindent To construct a complete orthonormal basis, we follow the following 
angular momentum addition scheme: 
\bea
\vec{J}[1]+\vec{J}[2] \rightarrow \vec{J}[12]+\vec{J}[3] \rightarrow 
\vec{J}[123]+...\vec{J}[1..(2d-1)]+\vec{J}[2d] \rightarrow \vec{J}[12..(2d)] = 0. 
\label{scheme} 
\eea 
Note that (\ref{scheme}) is a 2d-1 step process and the last step implies that the eigenvalues of 
$\vec{J}[12..(2d-1)]$ and $\vec{J}[2d]$ are equal. What follows now is an appropriate
generalization of the technique developed in \cite{schwinger}.  
To add the 2d angular momenta above, we consider direct product of the generating functions of 
two SU(2) coherent states defined over the complex planes x $(\equiv (x_1,x_2))$ and y $(\equiv (y_1,y_2))$ 
respectively. 
\bea 
\vert x \rangle \otimes \vert y \rangle \equiv  
\sum_{j_{1}m_{1}j_{2}m_{2}} \phi_{j_{1}m_{1}}(x)\phi_{j_{2}m_{2}}(y) 
\vert j_{1}, m_{1} \rangle \otimes \vert j_{2}, m_{2} \rangle 
= exp\left(x.a^{\dagger}[1] \oplus y.a^{\dagger}[2]\right) \vert 0 \rangle  
\label{id} 
\eea
where, 
\bea
\phi_{jm}(x) \equiv { \left(x_{1}\right)^{j+m}  \left(x_{2}\right)^{j-m} 
\over \sqrt{(j+m)!(j-m)!}} \nonumber 
\eea
We apply the differential operator involving a triplet of complex parameters 
$(\delta_1,\delta_2,\delta_3)$ and a complex plane $z (\equiv (z_{1},z_{2}))$: 
\bea 
exp\left(\delta_{3}~(\partial_{x}.\tilde{\partial}_{y})  + \delta_{1}~(z.\partial_{x}) + 
\delta_{2}~ (z.\partial_{y}) \right) 
\label{si} 
\eea  
on the both sides of (\ref{id}) and put $x = y =0$ to get \cite{schwinger}\footnote{Infact, 
the relation (\ref{12}) is a simple consequence of the fact that it's right hand side 
is a generator of SU(2) coherent states in the representation spaces of the total angular 
momentum operators.}: 
\bea
\sum_{j_{1}j_{2}jm} 
\phi_{jm}(z)
\Phi_{j_1 j_2 j} (\vec{\delta}) \vert|j_1 j_2 j m \rangle 
 =  exp\Big(\delta_{3}~a^{\dagger}[1].\tilde{a}^{\dagger}[2] + z.a^{\dagger}[12]\Big) 
\vert 0 \rangle  
\label{12} 
\eea 
where,
\bea 
a^{\dagger}_{\alpha}[12]   \equiv   \delta_{1} a^{\dagger}_{\alpha}[1] + 
\delta_{2} a^{\dagger}_{\alpha}[2] 
\label{idj}
\eea
\bea 
\Phi_{j_1 j_2 j} (\vec{\delta}) \equiv  \left[{(j_1 + j_2 + j+1)!  \over (2j+1)}\right]^{1 \over 2}   
{(\delta_{1})^{j_1 - j_2 + j} (\delta_{2})^{-j_1 + j_2 + j}   (\delta_{3})^{j_1 + j_2 - j} 
\over \left[(j_1 - j_2 + j)! (-j_1 + j_2 + j)! (j_1 + j_2 - j)!\right]^{1 \over 2}}~~.       \nonumber  
\eea 
For later convenience, $\Phi_{j_1 j_2 j} (\vec{\delta})$ will be 
called vertex factor  as they 
are involved in adding the  angular momenta $j_{1}$ and $j_{2}$ to give total angular momentum j
corresponding to the first step of the scheme (\ref{scheme}). If we now put:
\bea 
\vec{\delta} = (0,0,\delta_3)~~  => ~~  j_1 = j_2, ~~ j=0. \nonumber 
\eea 
Therefore, in this (trivial) d=1 case, the manifestly SU(2) gauge invariant Hilbert space at a 
particular site is: 
\bea 
\vert j_1 =j, j_2=j; j_{total}=0 \rangle = {1 \over \sqrt{(2j)!(2j+1)!}} 
\big(a^{\dagger}[1].\tilde{a}^{\dagger}[2]\big)^{2j} \vert 0 \rangle .  
\label{d1} 
\eea
Note that the states in (\ref{d1}) form a complete set of orthonormal and manifestly gauge 
invariant basis and  the U(1) Gauss law (\ref{consho}) makes j site independent. 
These results are obvious to begin with and did not require any calculations. 
We now generalize this procedure to arbitrary d dimensions.  To do this, we define 
(2d-1) triplets of complex parameters $\vec{\delta}[r]$ and the corresponding vertex  factor 
$\Phi_{j_{12..r}j_{r+1}j_{12..(r+1)}}(\vec{\delta}[r])$ associated with adding the angular momenta 
$j_{12..r}$ to $j_{r+1}$ to give net angular momentum $j_{12..(r+1)}$. This is the $r^{th}$ step of 
the ladder (\ref{scheme}). Therefore,  iterating the above process sequentially at all the 
2d-1 steps of the ladder (\ref{scheme}), we  get: 
\bea 
\sum_{\vec{j},m} \Big[\prod_{r=1}^{2d-1} \Phi_{j_{12..r}j_{r+1}j_{12..(r+1)}}(\vec{\delta}[r])\Big]  
\phi_{jm}(z) \vert j_1 j_2 j_{12} j_3 j_{123}..;j \equiv j_{12...2d}~ m \rangle \nonumber \\ 
=exp\big(z.a^{\dagger}[12..2d]\big) 
exp \sum_{r=1}^{2d-1} \delta_{3}[r] \big(a^{\dagger}[12..r].\tilde{a}^{\dagger}[r+1]\big) \vert 0 \rangle .  
\label{gfff}
\eea
In (\ref{gfff}), the summation over ${\vec{j}}$ is the summation over all (4d-1) angular momenta shown 
in  (\ref{scheme}) and the summation over m is from -j to + j. The generalization of (\ref{idj}) is: 
\bea
a^{\dagger}_{\alpha}[12..r]  \equiv  \delta_{1}[r-1]a^{\dagger}_{\alpha}[12..r-1] 
+  \delta_{2}[r-1]a^{\dagger}_{\alpha}[r].   
\label{no}
\eea
Further, like in d=1 case, to get the generating function of the gauge invariant basis we put:
\bea 
\vec{\delta}[2d-1] = (0,0,\delta_{3}[2d-1])~~ => ~~  j_{12..(2d-1)}= j_{2d}, ~~ 
j = j_{12...2d} =0.  \nonumber 
\eea
This finally gives the generating function of a manifestly SU(2) gauge invariant basis: 
\bea 
\sum_{\vec{j}} \Big[\prod_{r=1}^{2d-1} \Phi_{j_{12..r}j_{r+1}j_{12..(r+1)}}(\vec{\delta}[r])\Big]
 \vert j_1 j_2 j_{12} j_3 j_{123} j_4 ..; j = 0 \rangle  \nonumber \\
= exp \sum_{r=1}^{2d-1}\delta_{3}[r] \big(a^{\dagger}[12..r].\tilde{a}^{\dagger}[r+1]\big) \vert0\rangle .  
\label{gff}
\eea
To solve (\ref{gff}) for the physical states, we note that $(j_{12..r}-j_{r+1}+j_{12..r+1}), 
(-j_{12..r}+j_{r+1}+j_{12..r+1}), (j_{12..r}+j_{r+1}-j_{12..r+1})$ are all non-negative integers
and put 
\bea 
{\delta}_{1}[r] = exp i \theta_{1}[r], ~~ 
{\delta}_{2}[r] = exp i \theta_{2}[r] ~~ r=1,2....,2d-1.  
\nonumber 
\eea
This gives: 
\bea 
 \vert j_1 j_2 j_{12} j_3 j_{123} ...; j = 0 \rangle  = N \left[\prod_{r=1}^{2d-1}  \prod_{\alpha=1}^{2} 
{1 \over 2\pi} \int_{0}^{2 \pi}  d\theta_{\alpha}[r] 
\Phi_{j_{12..r}j_{r+1}j_{12..(r+1)}}(\vec{\delta}^{*}[r])\right] \vert 0 \rangle . 
\label{giob} 
\eea
In (\ref{giob}), ${\delta}_{\alpha}^{*}[r] =  exp -i \theta_{\alpha}[r],  \alpha=1,2$ 
and ${\delta}_{3}^{*}[r] \equiv a^{\dagger}[12..r].\tilde{a}^{\dagger}[r+1]$ are 
the manifestly SU(2) gauge invariant operators. Further, ${\delta}_{3}^{*}[2d-1]$ is 
independent of $\delta_{1}[2d-1], \delta_{2}[2d-1]$ and therefore, the corresponding 
integrations project out the total angular momentum zero states. N is the normalization 
constant at site n and is given by: 
\bea 
N \equiv \prod_{r=1}^{2d-1} { (2j_{12..(r+1)} + 1)! (j_{12..r}-j_{r+1}+j_{12...(r+1)})!
(-j_{12..r}+j_{r+1}+j_{12...(r+1)})!  \over {(j_{12..r}+j_{r+1}+j_{12...(r+1)} + 1)!}} . \nonumber  
\eea 
Therefore, at this stage the problem 
of solving the non-abelian Gauss law reduces to the problem of solving the abelian 
Gauss law (\ref{consho}).  This abelian Gauss law simply states $j(n,i)
= j(n+i,d+i)$  for i=1,2,..,d.
The complete $SU(2) \otimes U(1)$ invariant Hilbert space can be written as: 
\bea 
{\tilde{\cal H}}^{p}  = \prod_{n}{}^{\prime} \otimes {\tilde{\cal H}}_{n}^{p}
\label{phss} 
\eea 
In (\ref{phss}), the direct product is taken over all the lattice sites such that 
U(1) Gauss law (\ref{consho}) is satisfied.  Note that 1) the states in 
(\ref{giob}) are the eigenstates of the complete set of commuting angular momentum 
operators shown in (\ref{scheme}). Therefore, by construction they form orthonormal basis. 
2) they are manifestly SU(2) gauge invariant. As is clear from (\ref{giob}), we require 
(4d-3) quantum numbers at a given lattice site to 
characterize the SU(2) gauge invariant Hilbert space. Further, the U(1) Gauss law fixes 
d of these quantum numbers in terms of the quantum numbers associated with the previous 
sites.  Therefore, the $SU(2) \otimes U(1)$ 
gauge invariant Hilbert space is characterized by 3(d-1) quantum numbers per lattice site 
which is the number of physical transverse degrees of freedom of SU(2) gluons.  

We now briefly discuss the SU(N) lattice gauge theories. The SU(2) Schwinger boson 
algebra was generalized to SU(N) group in \cite{manu1}. This required introduction 
of (N-1) SU(N) fundamental multiplets of harmonic oscillators.  Therefore, the prepotential 
formulation of SU(N) lattice  gauge theory will be invariant under $SU(N) \otimes (U(1))^{N-1}$ 
gauge group. Like in the present case, this formulation is also useful in the construction 
of a manifestly gauge invariant orthonormal basis. The work in this direction is in progress 
and will be reported elsewhere \cite{manu2}.

{\section {Discussion and Summary}} 

\noindent 
In this work, we have constructed an orthonormal basis in the physical Hilbert 
space of SU(2) lattice gauge theories in arbitrary dimension. 
This construction unlike the earlier approaches has both the desired features: 
a) it is manifestly gauge invariant like Wilson loops basis b) it is orthonormal 
and complete like basis given in terms of Wigner D matrices.  
This is the first application of the harmonic oscillator prepotential formulation of lattice 
gauge theories.  The matter can be included as in \cite{manu}. 
Breaking the link operator U into left and right harmonic 
oscillator doublets was essential for this construction and it completely bypassed 
the permutation symmetry problems faced earlier.  Using (\ref{dh}), it 
would be interesting to study the connection of (\ref{giob}) with the Wilson loop 
overcomplete basis as both are manifestly gauge invariant.  

\noindent The investigation of lattice Hamiltonian eigenvalues and eigenvectors 
in the physical Hilbert space is an alternative to path integral formulation. 
It is generally the truncated basis with respect to the number of plaquettes which 
is included in such calculations.  Therefore,  the present work is going to be 
useful in this direction.  The matrix elements of the Hamiltonian and the 
corresponding dynamical issues are under investigation and will be reported 
at a later stage.

\end{document}